\documentclass{article}

\usepackage{arxiv}

\usepackage[utf8]{inputenc} 
\usepackage[T1]{fontenc}    
\usepackage{hyperref}       
\usepackage{url}            
\usepackage{booktabs}       
\usepackage{amsfonts}       
\usepackage{nicefrac}       
\usepackage{microtype}      
\usepackage{lipsum}
\usepackage{graphicx}
\graphicspath{ {./images/} }

\title{The Convergence of Blockchain Technology and Islamic Economics: Decentralized Solutions for Shariah-Compliant Finance}

\author{
 Naseem Alsadi \\
 McMaster University \\
 \texttt{alsadin@mcmaster.ca} \\
}

\begin{document}
\maketitle
\begin{abstract}
This paper provides a brief overview of the ongoing financial revolution, which extends beyond the emergence of cryptocurrencies as a digital medium of exchange. At its core, this revolution is driven by a paradigm shift rooted in the technological advancements of blockchain and the foundational principles of Islamic economics. Together, these elements offer a transformative framework that challenges traditional financial systems, emphasizing transparency, equity, and decentralized governance. The paper highlights the implications of this shift and its potential to reshape the global economic landscape.

\end{abstract}


\section{Blockchain and Money}

All things in society have an attributed cost and value, even money. The cost of money is called interest. The erosion of money’s value is called inflation. The relationship between the two is central to today’s economy. Centralized institutions, such as banks, will use the relationship between inflation and interest to influence the state of the economy. Take for instance an example of excessive monetary supply growth (centralized institutions increasing the amount of money circulating) leading to an increase in inflation. To combat this, the same centralized institutions will increase interest rates to reduce inflation. When interest rates rise, borrowing becomes more expensive. This discourages people and businesses from taking out loans to spend or invest. With less money circulating in the economy, demand for goods and services decreases, this helps slow down price increases, thus reducing inflation.

If you’re new here, you may be wondering “wait what do you mean print more money?” Let’s quickly review the history of this uncomfortable statement. The gold standard was a monetary system where a country’s currency was directly linked to a finite amount of gold. Centralized institutions could not print money at will but rather had to acquire more gold first. This limited natural money growth. However, the United States stopped using the gold standard in 1933, ultimately cutting all ties to the system in 1973. In return, the gold standard was replaced with the fiat system. Fiat money is a currency, which unlike its predecessor, is not backed by precious finite metals and therein possesses no intrinsic value except the value that users agree on. The key here being the lack of backing by a finite resource. Meaning that centralized institutions can print more at will. As a result, as we have seen, excessive monetary supply growth has increased inflation, making it easier to borrow cheaply, even for risky ventures. As a result, risk may be underpriced — loans or investments may carry less cost than they should, given the actual likelihood of failure or loss.

Returning back to the initial point at hand. Blockchain provides the solution and this solution converges to Islamic economics. Similar to the gold standard, limited supply and mining difficulty build intrinsic value into Bitcoin by ensuring that it cannot be easily inflated or manipulated by centralized institutions. Just as gold’s scarcity and the effort required to extract it contribute to its value, Bitcoin’s capped supply of 21 million and the increasing difficulty of mining make it a deflationary asset. Take for example Bitcoin, proposed in 2009, which has a maximum supply capped at 21 million coins. Given the number of Bitcoins mined so far $M$, Bitcoin’s scarcity $S$ is calculated as the ratio of the remaining supply to the coins already mined:
\[
S = \frac{21,000,000 - M}{M}
\] 

This is in contrast to modern financial systems, which chase low-volatility capital appreciation. With Bitcoin, the market would naturally price risks more appropriately, which would lead to fewer speculative debt contracts, loans or credit agreements where borrowers take on debt, often with the expectation that future economic growth will allow them to repay the loans with interest. Instead, there would be a shift toward equity-based, risk sharing financial models. Equity financing aligns the interests of investors and borrowers more closely, as they both share in the risk and rewards of an enterprise, rather than creating fixed obligations like interest payments. Investors are not guaranteed fixed returns, but instead are directly tied to the performance of the business or investment. Since Bitcoin’s supply can’t be manipulated, the real value of assets becomes clearer. There is less artificial inflation, so businesses and investors can make decisions based on actual value rather than speculative growth. Investors can’t rely on centralized institutions to interfere and make their debts easier to repay over time. Instead, they must carefully assess the real risks of an investment. If the business is risky, that risk will be reflected in the investment terms.

\section{Islamic Finance}

This is precisely the logic that Islamic finance employs. Risk-sharing in Islamic finance aligns with ethical imperatives derived from Shariah, particularly the prohibition of riba (interest) and gharar (excessive uncertainty). By sharing risks, both investors and entrepreneurs are incentivized to manage the venture prudently, ensuring that neither party unfairly profits from the misfortune or exploitation of the other. Investors and entrepreneurs share profits and losses according to pre-agreed ratios, fostering a sense of shared responsibility and fairness. Two primary contracts, Musharakah (joint partnership) and Mudarabah (profit-sharing), exemplify this principle. In a Musharakah contract, all parties contribute capital and share in profits or losses based on their capital contribution. In a Mudarabah contract, one party provides the capital while the other provides the management expertise, with profits shared based on a predetermined ratio, though losses are borne solely by the capital provider. This model stands in contrast to interest-based lending, where the lender earns a fixed return irrespective of the borrower’s outcome, transferring the entire financial burden and risk to the borrower. Islamic economic theory argues that risk-sharing not only enhances financial justice but also fosters a more stable economic system by reducing the tendency toward speculative bubbles and excessive debt accumulation, as seen in conventional risk-transfer models.

\section{Blockchain and Islamic Finance}
\subsection{Musharakah}
For Musharakah, blockchain serves as an ideal infrastructure to facilitate joint ventures where participants pool resources and share profits or losses based on pre-agreed ratios. Contributions to these ventures can be tokenized, representing proportional ownership. These tokens, governed by smart contracts, ensure that profits are distributed transparently and automatically in line with the agreed terms, eliminating the need for intermediaries and manual reconciliation. The immutable ledger of blockchain provides an additional layer of security by recording all transactions, contributions, and financial outcomes, allowing all parties to verify their stakes and returns in real time. This ensures that the ethical requirements of Musharakah, such as fairness and transparency, are met while addressing potential challenges like misrepresentation or disputes. Blockchain also enhances Musharakah by introducing liquidity and flexibility to partnerships. Tokenized ownership allows stakeholders to transfer or sell their stakes without disrupting the underlying venture. This enables participants to adjust their investment positions based on their financial needs while preserving the integrity of the Musharakah agreement. Such features make blockchain-based Musharakah ventures more dynamic and accessible, particularly for smaller investors who might not otherwise have the means to participate in large-scale partnerships. Additionally, the decentralized nature of blockchain ensures that decision-making processes within the partnership remain equitable. Governance mechanisms, built into the blockchain infrastructure, allow all stakeholders to vote on important decisions, fostering shared responsibility and collective accountability.

\subsubsection{Mudarabah}

Similarly, blockchain technology provides the foundation to address the operational challenges inherent in Mudarabah. The relationship between the capital provider (rabb-ul-mal) and the entrepreneur (mudarib) in Mudarabah is predicated on trust and accurate reporting of profits and losses. Blockchain eliminates the need for blind trust by providing a transparent and immutable ledger of all financial activities and performance metrics. This shared record ensures that both parties have a clear and real-time view of the venture’s financial health, reducing information asymmetry and fostering confidence. Smart contracts further enhance Mudarabah by automating the profit-sharing process. Once the venture generates revenue, these contracts execute the distribution of profits based on the pre-agreed ratios, ensuring that the terms are upheld without delay or ambiguity. Blockchain also introduces scalability to Mudarabah arrangements. By tokenizing capital contributions, blockchain enables multiple investors to pool their resources into a single venture. Each investor holds a proportional share of the venture, represented by tokens that entitle them to a corresponding share of the profits. This democratizes access to Shariah-compliant financial opportunities, enabling smaller investors to participate in ventures that would traditionally require substantial capital. Entrepreneurs, in turn, benefit from a diversified pool of investors, reducing their reliance on a single source of funding and distributing risk more equitably. Blockchain’s decentralized structure ensures that all investors are treated fairly and that their contributions and returns are transparently recorded. Moreover, blockchain mitigates disputes and inefficiencies often associated with conventional financial systems. The automation provided by smart contracts ensures that profit distribution is instantaneous and compliant with the agreed terms. Losses, too, are recorded transparently, allowing the rabb-ul-mal to verify that they are not the result of negligence or misconduct by the mudarib. Blockchain’s immutable audit trails provide a reliable mechanism to attribute losses fairly, upholding the principle that capital providers bear financial losses unless proven otherwise. This level of accountability discourages unethical behavior and promotes prudent management by the mudarib, aligning with the ethical imperatives of Islamic finance. In addition to these functional benefits, blockchain introduces innovative governance mechanisms for Musharakah and Mudarabah ventures. Stakeholders in Musharakah agreements, for example, can participate in decentralized decision-making processes, voting on critical matters such as reinvestment strategies or business expansions. This ensures that all parties have an equitable voice in the partnership, fostering shared responsibility and mutual accountability. In Mudarabah, the performance-based rewards for the mudarib can be encoded into smart contracts, tying their returns directly to the venture’s success. This promotes prudent management and discourages speculative behavior, aligning with the risk-sharing ethos of Islamic finance.

\section{Conclusion}

Blockchain technology has emerged as a transformative force, redefining traditional financial systems through its decentralized, transparent, and trustless nature. In the context of Islamic finance, these features provide a groundbreaking avenue for implementing Musharakah and Mudarabah, two cornerstone principles that emphasize risk-sharing, ethical collaboration, and equitable profit distribution. By leveraging blockchain, these contracts are no longer constrained by conventional limitations such as geographic boundaries, operational inefficiencies, and challenges in trust establishment between parties. The scalability of blockchain ensures that Musharakah and Mudarabah agreements can operate on a global scale, engaging participants from diverse regions with minimal friction. Smart contracts—self-executing agreements embedded in the blockchain—eliminate ambiguities, enforce compliance with Shariah principles, and automate processes such as profit-sharing and liquidation. These attributes not only enhance operational efficiency but also build trust among stakeholders, ensuring that the ethical integrity of Islamic finance remains intact. Moreover, blockchain democratizes access to financial instruments by lowering entry barriers. Small and medium-sized enterprises, often excluded from conventional financing due to high costs or lack of collateral, can benefit from transparent and cost-effective blockchain-powered Musharakah and Mudarabah contracts. This inclusivity fosters economic empowerment and aligns with the ethical objective of Islamic finance to promote social justice and shared prosperity. From an ethical perspective, blockchain aligns seamlessly with the principles of Maqasid al-Shariah (the objectives of Islamic law), which emphasize transparency, fairness, and the avoidance of exploitation. The immutable nature of blockchain records ensures that all transactions are transparent, reducing the risk of disputes and unethical practices. This trustless environment mitigates human error and corruption, reinforcing the moral and ethical foundation of Islamic financial practices. Ultimately, blockchain transcends its role as a technological tool, becoming a vehicle for manifesting the core values of Islamic finance. It provides the infrastructure to actualize principles of justice and equity in a practical and scalable manner. By integrating blockchain into Islamic finance, a more inclusive, ethical, and innovative financial ecosystem can emerge—one that is not only transformative for the Islamic finance sector but also serves as a model for ethical financial systems worldwide. The synergy between blockchain and Islamic finance opens new horizons, proving that technological innovation can coexist with, and even amplify, ethical and economic ideals.



\end{document}